# High temperature mediated rocksalt to wurtzite phase transformation in cadmium oxide nano-sheets and their theoretical evidence

**Arkaprava Das[1]\*, C. P. Saini[1], Deobrat Singh[2]\*, Rajeev Ahuja[2] and Sergei Aliukov[3]**

[1]Inter University Accelerator Centre, Aruna Asaf Ali Marg, New Delhi-110067, India
[2]Department of Physics and Astronomy, Condensed Matter Theory Group, Uppsala University, Sweden
[3]South Ural state University, Chelyabinsk, Russia

## Abstract

In the paper, high temperature induced phase transformation (PT) in chemically grown CdO thin films has been demonstrated whereas their corresponding electronic origin is further investigated by density functional theory. In particular, cubic rocksalt to hexagonal wurtzite PT in 900 ℃ annealed CdO thin films is confirmed by X-ray diffraction (XRD), consistent with High Resolution Transmission Electron Microscopy (TEM). Such high temperature treatment also leads to significant enhancement in optical band gap from 2.2 to 3.2 eV as manifested by UV-Visible spectroscopy. Moreover, atomic force microscopy and scanning electron microscopy clearly evidence the structural evolution via formation of nano-sheet network in wurtzite phased CdO films. Furthermore, X-ray Absorption spectra at oxygen *k*-edge revealed a notable shift in inflection point of absorption edge while X-ray Photoelectron spectroscopy of Cd 3d and O 1s suggested the gradual reduction in $CdO_2$ phase with increasing annealing temperature. In addition, different complementary techniques including Rutherford Backscattering, Raman Spectroscopy have also been exploited to understand the aforementioned PT and their structural correlation. Finally, molecular dynamics simulation along with density functional theory calculations suggest that symmetry modification at Brillouin zone boundary provides a succinct signature for such PT in CdO thin film.

**Key words**:CdO Nanosheets; Cubic Rocksalt; Hexagonal Wurtzite.

*Author for correspondence:* arkaprava@iuac.res.in (Dr. Arkaprava Das, Research Scholar)

## I. INTRODUCTION

Cadmium Oxide (CdO) is an inherently *n*-type degenerate semiconductor. It is known that predominantly due to the existence of Cd interstitial $Cd_{(1+x)}O$ or oxygen vacancies $CdO_{(1-x)}$ in the cubic lattice, its direct band gap changes from 2.2 to 2.6 eV and its indirect band gap varies from 1.36 -1.98 eV [1].CdO is also having a high exciton binding energy of 75 meV[2]. It crystallizes in rocksalt phase and shows transparency almost in the whole visible region of the electromagnetic spectrum [3]. In the recent few decades, CdO has enticed the interest of scientific community for its distinguishable features like high intrinsic mobility, high carrier concentration without any external doping, high transparency etc. Therefore, CdO becomes potentially an ideal candidate to be utilized as transparent conducting oxide (TCO) for optoelectronic devices operating at lower wavelengths [4] region, flat panel display [5] and thin film photovoltaic. It is also having applications as photodiodes, photo transistors, transparent electrodes,[6]gas sensors, catalysts, [7] IR detectors and liquid crystal display [8,9]. Various types of CdO nanostructure morphologies have been reported in the literature to date, such as nanowires, [10]nanotubes, [11]nanoparticles, [12]nanofilms, [13]nanoneedles, [8] and nanorods[14] etc. Diverse techniques has been used in the literature for CdO thin film preparation such as chemical vapour deposition (CVD), [13]spray pyrolysis, [15] pulsed laser deposition, [16] chemical bath deposition (CBD)[17] etc.

In the present investigation, synthesis of solgel prepared CdO thin films on Silicon and Silica substrate and their post annealing at 500, 700, 900°C in oxygen environment have been accomplished. In this manuscript, growth mechanism of nanocrystallites and nanosheet networks, and its correlation with cubic rocksalt to hexagonal wurtzite phase transformation at 900 °Chas been discussed in corroboration with structural, microscopic and optical properties. To detect the phase transformation in thin films, observation of band gap modification would be a trustworthy idea [18]. The 2D nanosheet networks and hexagonal shaped nanodiscs are advised to be convenient for optoelectronic applications such as data storage, memory devices, energy storage and energy conversion [19]. It has been reported by Umar *et al.*[20] that in case of ZnO, nanosheet network formation takes place at the same hexagonal wurtzite phase at around 650 to 700°C but interestingly in case of CdO the same is found to take place accompanied with a structural phase transformation. However, such self-assembled nanostructure might provide a distinguishable root for betterment of physical properties of an anisotropic material due to their enhanced rough surface structure made at the cost of nanoparticles[21]. Generally nanocrystallites growth takes place via Ostwald ripening [22] or oriented attachment process [23]. However, cluster migration process can't be ruled out completely[24]. In Ostwald ripening process, crystal growth takes place at the cost of smaller ones having a surface energy difference and in oriented attachment process adjacent particles shares a common crystallographic orientation and get self-assembled. Formation of nanosheet like unique structure is the consequence of growth kinetics. There are two kind of growth mechanism for nanowires firstly vapour-liquid-solid (VLS)[25] process secondly vapour-solid (VS)[26] process. In the first one, metal catalyst is used at the growth front. In the second growth, process oxide evaporates from higher temperature region and sediments on a lower temperature region on the substrate. In the present investigation nanocrystallites and nanosheet formation are anticipated by Ostwald ripening and VS process. The correlation of band gap enhancement with structural phase transformation has been explained in this work. Further, analysis of oxygen $k$ edge and the shifting of inflection point of absorption edge is well correlated with band gap enhancement. X-ray Photoelectron Spectroscopy (XPS) measurements indicate the existence of contamination phases which get significantly modified with Ar ion etching. $CdO_2$ surface phase,[27] gets disappeared completely for CdO900 thin film which reflects its explicit dependence upon thickness. Density functional theory (DFT) and molecular dynamics (MD) simulation reflects that lattice relaxes to wurtzite phase for CdO900 thin film and critical analysis of band structure reflects that valence band maxima (VBM) shifts from zone boundary to centre of the Brillouin zone (BZ) i.e. at Γ point for transformed wurtzite with a change in the direct as well as indirect band gap[28]. Such kind of temperature driven PT with theoretical evidence is very rare in the literature to the best of our knowledge. Therefore, the present manuscript describes the nanosheet formation and its deep correlation with temperature dependent structural, optical and electronic properties which further supports the structural phase transformation phenomena. A complete experimental and theoretical study has been accomplished for understanding the PT.

## II. EXPERIMENTAL

Cadmium oxide thin films were deposited on silica and Si substrate using solgel spin coating method. Details of the sample preparation are mentioned elsewhere[29]. Further, the thin films are annealed in 500, 700, 900 °C at oxygen environment.

All the instrumental facilities utilized for all the characterizations, apart from TEM and XPS measurements are enunciated elsewhere[29].

Analysis regarding the contamination phases and other existing surface phases has been performed by XPS. The concerned instrument belongs to Omricon nanotechnology XPS system from Oxford instruments with ESCA+ model specification. The system is consisted of an ultra high vacuum chamber connected with Al-$K_\alpha$ source having energy 1486.7 eV and a 124-mm hemispherical electron analyzer. Surface etching has been performed with 6 KeV Ar ion inside the chamber. Nanoparticles diameter, inter-planner distance for the thin films have been investigated using HRTEM, model, Tachno FEI with 300 kV operating accelerating voltage. Sample preparation has been accomplished by peeling of the thin film from the substrate and subsequent bath sonication process.

## II.1. COMPUTATIONAL METHODS

The electronic structure calculations are based on density functional theory (DFT) and have been performed using Vienna Ab initio Simulation Package (VASP) software [30–32]. The local density approximation (LDA)[33,34] exchange correlation functional along with projector augmented wave (PAW) method[35,36] was employed in the VASP software. An energy cut-off of 1200 eV has been used for the plane-wave basis set and 15×15×15 set of special k-points sampling for the integrations over the BZ[37] has been used for achieving the accurate electronic structures for the rocksalt and wurtzite phase of CdO. Generally, LDA gives lattice parameter within 2% of the experimental values, but it rigorously underestimates the electronic band gaps. For semiconductors which have semi-core cation d states, the error is mainly severe because the LDA gives underestimated binding energy in case of d-states and consequently, overestimating their hybridization with anion p-states and increasing the effect of p-d coupling. In this paper, we have used the LDA+U methods[38,39] for on-site Coulomb interaction $U_{eff}$ = U−J, where U is the Hubbard parameter and J is the exchange parameter. During the electronic structure calculations, we utilized U = 12.0 eV, and J = 2.10 eV for correlation effect of localized d-states for Cd atom. The full convergence criteria for energy in the self-consistent loops and maximum force tolerance were set to be $10^{-6}$ eV and 0.002eV/Å, respectively. The ab-initio molecular dynamics calculations were performed for electronic structure calculations in the micro-canonical ensemble in which volume, particle number and temperature (VNT) were fixed.

## III. RESULTS AND DISCUSSIONS

### III.1. Compositional studies

Compositional analysis for the three thin films annealed at 500, 700 and 900 ºC (henceforth, referred to as CdO500, CdO700 and CdO900, respectively) has been done using RBS technique. The thicknesses and compositional percentages of these three thin films have been calculated using RUMP simulation and the values are cited in Table 1.Figure 1 shows the simulated RBS spectra for all the thin films. We observe that with increasing annealing temperature the film thickness decreases and it means that there is material loss at higher temperatures. The thickness of the film is observed to be lowest for CdO900. CdO melting temperature is in the range of 900 ℃-1000 ℃. Therefore, for CdO900 thin film, CdO might have melted and after condensation, it forms a nanosheet like network throughout the substrate which is quite clear from the SEM, AFM and TEM images described in the upcoming section of the manuscript. So it is expected that the overall thickness of CdO900 thin film is less as compared to the CdO700 and CdO500 thin films. It is also observed that the yield of Cd has also reduced drastically for CdO900 thin films since at 900 ℃ annealing temperature CdO starts to melt and flowed out of the substrate during annealing. Therefore, the percentage amount of Cd reduced significantly. Depth profiling plots have been shown at figure 1 in the supplementary datasheet. However, the reduction is more prominent for Cd compared to the O. The linear diffusion of Cd inside Si layer at the interface region provided for fitting also increases with increasing annealing temperature. Therefore, the drastic change in thickness and stoichiometry for CdO900 gives an indirect signature that there might be a possibility of phase transformation.

**III.2. Structural properties**

Figure 2 shows the glancing angle diffraction patterns for CdO500, CdO700, CdO900 thin films. From the XRD patterns, we found that the CdO500 and CdO700 thin films have polycrystalline nature with a cubic structure (JCPDS: 78–0653) at room temperature but in CdO900 thin film polycrystalline nature gets vanished completely which is clearly observed from the microscopy images in the upcoming section. Therefore, the XRD pattern also completely changes for that thin film. The diffraction peak intensity corresponding to (111) peak reduced for CdO700 thin film compared to CdO500 one. For CdO900 thin film, not only (111) peak but also (200) and (220) peaks get vanished completely and sharp tiny peaks evolved at 37.1 degree, 45.9 degree, 48.6 degree, 54.8 degree, 56.8 degree are generated. Among them, peak situated at 37.1 and 48.6 degree provide the signature of (101) and (102) plane for hexagonal wurtzite phase. The particle diameter has been calculated using Scherrer formula considering the FWHM of (111) peak. The formula is given by the following equation [40]

$$D = \frac{k*\lambda}{\beta \cos\theta} \quad (1)$$

Where, $\lambda$ is the X-ray wavelength, $\theta$ is the Bragg's diffraction angle (half of the peak position angle), $\beta$ is the full width half maximum of the main peak in XRD pattern in radian.

K is the shape factor whose value is generally taken as 0.94. The calculated values of particle diameter and inter planner spacing are cited in Table 1 in supplementary data sheet.

The particle diameter enhanced in CdO700 thin film which gives a direct signature of enhanced crystallinity. However the intensities of all the peaks decreases but FWHM of (111) peak reduced little bit. Therefore, small enhancement of particle diameter of takes place. There is no change in peak position or 2θ values for all (111), (200), (220) peaks for CdO500 and CdO700 thin films which reveals that there is no compressional or tensile stress generation due to annealing at 700 ℃. At 900℃, we found nanosheet formation in the present case. It is reported by A. Umar *et al.*[20] that in case of ZnO, nanosheet formation also takes place at higher annealing temperature and peaks in the XRD pattern gets sharper which gives a signature that nanosheet formation takes place in the same wurtzite phase. However, in our case, from the completely changed XRD pattern, an indirect remark can be made that the nanosheet formation is not taking place in the same cubic structure for CdO900 thin film but in wurtzite phase.

**III.3. Microscopy analysis**

Figures 3 (a), 3(b), 3(c) show the SEM images of CdO500, CdO700, CdO900 thin films, respectively and figure 3(a'), 3(b'), 3(c') are higher magnification imaging of the same. Figure 4(a), 4(a') shows the EDX spectra for CdO900 thin film. From SEM images, we observe that CdO500 thin film is feebly crystallized as compared to the CdO700 thin film without any voids. Figure 5(a), 5(b), 5(c) represents the AFM images for CdO500, CdO700, CdO900 thin films in 5 μm scale whereas figure 5(a'), 5(b'), 5(c') represents the same in 2 μm scale. All SEM and AFM images reflect that the films are quite uniform without any huge crack or dust particle. On correlating with XRD results, it can be concluded that the structures seen in these images are aggregation of several small size crystalline grains. The increment in average grain size reduces the overall grain boundary area and subsequently reduces overall grain boundary scattering. Usually nanocrystallites formation takes place via nuclei formation, coalescence and growth process. From lower annealing temperature coalescence process gets initiated by small crystallites as it starts to get kinetic energy. Generally for annealing below 500℃, cluster migration process predominates to initiate nanocrystallites formation. In the present case, annealing temperature is above 500℃. So here the predominating process for the formation of nanocrystallites would be Ostwald ripening process[41] which is a thermal energy dependent process and becomes more significant as the annealing temperature rises[24]. Generally, in Ostwald ripening process larger crystallites formation occurs at the cost of smaller crystallites because of energy difference between larger particles and smaller particles of a higher solubility based on Gibbs-Thompson rule[42]. Usually at higher calcinations temperature atoms inside the system get sufficient energy to readjust their positions and lattice tends to get relaxed. From both SEM and AFM images, it is clear that the nanocrystallites size enhanced for CdO700 compared to CdO500 thin films via Ostwald ripening which is undoubtedly compatible with XRD results.

Figure 6(a), 6(a') shows the AFM and SEM images of CdO900 thin film. From figure 6, we can observe that nanosheet network structure has formed throughout the substrate. For CdO900 thin films, we can't see the nanocrystallites anymore. Instead of expected larger nanocrystallites, we observe a completely changed morphology, a nanosheet like structure. The particle like features which we have found quite prominently in CdO700 thin film gets disappeared from the surface. In section analysis (shown in the supplementary data sheet, figure 2) from nanoscope software we found the height scale as well as the relative height between peaks and valleys are different for the three thin films. For CdO700 thin film, it seems that nanocrystallites expanded in x-y plane at the cost of z axis which means nanocrystallites growth has been taken place parallel to the substrate surface at the cost of vertical growth. From this observation one can anticipate that possibly a sufficient amount of force develops at the grain boundaries to curl the z axis growth vertical to the substrate surface for CdO700 thin film. But in case of CdO900, surface becomes quite rough compared to the others which can be clearly observed from the rms roughness value. The values of *rms* roughness calculated from nanoscope software are cited in table 2. Now this drastically changed morphology, nanosheet formation for CdO900 thin film has deep correlation with CdO cubic rocksalt to hexagonal wurtzite phase transformation which has been discussed in the next section.

Annealing has a significant effect on the surface features as it affects the surface migration properties and surface energies which might lead to formation of new surface structures and crystallites consisting different size and nature which is exactly taking place at the present case at different annealing temperature. It is a well-established phenomenon that growth of nanosheet and ribbon like nanostructure is mainly dominated by vapour-solid (VS) process [26]. In this process oxide vapour at higher temperature evaporates from the initial oxide at a higher temperature zone and directly deposit on the substrate surface where the temperature is lower. Usually more relaxed and defect free nanocrystallites becomes more reactive [43]. In the present case, with increasing annealing temperature the increased reactivity would certainly increases nucleation probability for formation of nanosheet like structure for CdO900 thin film. Here, at annealing temperature of 900℃, which is nearer to the CdO melting temperature, the oxide gets melted and subsequently they condensate and nucleate at energetically favoured site on the substrate. It is reported for hexagonal wurtzite ZnO that it has a positively charged (0001) Zn terminated plane and negatively charged (000$\bar{1}$) O terminated plane between which (0001) Zn is chemically more active [44]. Interestingly from our energy dispersive X-ray spectroscopy (EDX) as shown in figure 4 for CdO900 thin film, we observe that the whitish portions are Cd rich and blackish or darker region are Cd deficient, relatively O rich. From this observation, one can anticipate that at 900 ℃ CdO starts to behave like a hexagonal wurtzite polar crystal where positively charged more active Cd (0001) planes further energetically favour the subsequent adsorption and nucleate such chemically active planes throughout the substrate which has been observed as Cd rich regions in SEM images and forms a nanosheet like structure. Therefore, it is obvious that thickness of those whitish regions would be more compared to the darker or blackish regions. However the enhancement in weight percentage is more prominent for Cd compared to O that's why those elevated whitish portions are said to be as Cd rich region. We know that the carrier concentration of CdO is very high in the order of $10^{19}$/cc. Due to this higher electron concentration, it can influence the structural phase. Since, the existing basal plane electrons endorses a field induced force which might help to align the dipole moments via spontaneous polarization induced electrostatic energy [45] not only for the minimization of these electrostatic energy generated via polar planes but also to balance the difference in surface tension of the formednanosheets[46]. These nanosheets are mingled together throughout the substrate mainly due to the aforementioned long range electrostatic interaction among the polar charges of the (0001) planes and form a nanosheet network[20]. Obviously under thermodynamic equilibrium, the higher surface energy facet occupies small area and lower surface energy facet occupies larger area for minimization of overall surface energy. Therefore, in the nanosheet it has been observed that Cd rich regions are less abundant compared to the blackish O rich region. However, the sheet has been formed at 900 ℃ not at 700 ℃ as for crystal step growth less energy is required compared to the secondary nucleation events at finite volume [47] and also the melting temperature of CdO is around 900 ℃-1000 ℃. Therefore, it can be remarked that with increasing annealing temperature rocksalt CdO nanocrystallites transforms into CdO nanosheet, accompanied by a parallel phase transformation from cubic rocksalt to hexagonal wurtzite phase which has been cross verified using the optical study which will be discussed in the upcoming section. So from the above discussions, it can be concluded that the nanosheet formation takes place mainly via VS process, however it is difficult to completely rule out vapour-liquid-solid (VLS) process [25]. Since due to low melting temperature, Cd particles might act as catalyst for the growth process also. A schematic illustration for formation of nanosheet has been shown at figure3S in the supplementary data sheet.

TEM was conducted for a comprehensive analysis of the shapes and sizes of CdO500 and CdO900 thin films. Figures 7(a) clearly shows the formation of almost spherical nanoparticles with different sizes in CdO500 thin films. The particle size of these nanoparticles has been calculated using Image J software and ranges from 6 to 8 nm which is much lesser than particle diameter calculated from Debye Scherrer formula from XRD pattern and that is possible as for TEM samples are prepared by peeling of thin films from substrate and subsequent bath sonication process[48]. In addition, the calculated value of lattice spacing from HRTEM patterns (Figure 7(b)) is found to be 0.239 nm corresponding to inter-planar distance of (200) plane (using 'IMAGE J' software) positioned at 38.4° in XRD pattern. On the other hand, as we inferred clearly from SEM and AFM images that there are changes in morphology and topography of the thin films with annealing from 500 ℃ to 900 ℃, the same can be observed in TEM images. From figure 7(c), we can clearly observe the formation of nanosheets in CdO900 thin films. The inter-planar spacing equal to 0.242 nm, calculated from the HRTEM image for CdO900 thin films (Figure 7(d)) matches well with that of plane (101) positioned at 37.1° in XRD pattern. Therefore, TEM analysis also endorses our XRD results and provides a direct experimental signature of rocksalt to wurtzite phase transformation.

### III.4. Optical properties

Optical transmittance spectra are shown in Figure 8(a) for CdO500, CdO700, CdO900 thin films on silica substrates and in the inset the absorption spectra of the all those three thin films have been shown. The spectra reveal the average transmittance for all the three thin films in between 45% to 60% within the wavelength range of 600 nm to 800 nm. However among these three thin films, CdO900 spectrum shows a lower percentage of average transmittance in that wavelength range. These transmittance spectra do not exhibit any interference fringes which can be understood by Pankov analysis[49]. There is not much significant change in near band edge absorption between 350 nm to 600 nm for both CdO500 and CdO700 thin film which is clearly observed from both the transmittance and absorption spectra. But there is a prominent blue shift of the cut-off wavelength for CdO900 thin film in transmittance as well as in absorption spectra which gives a direct signature of band gap enhancement. This sharp change is also clearly visible in Tauc plots which have been used for band gap determination. The band gap values are cited in table 1. Figure 8(b) shows the Tauc plot and in the inset specifically Tauc plot for CdO900 thin film has been shown. The optical band gap has been calculated using the following Tauc equation[50]

$$\alpha h\nu = A(h\nu - E_g)^n \quad (2)$$

In the above equation A is a constant, hν is the photon energy, $E_g$ is the optical band gap of the thin film to be calculated. Here the exponent n depends on the type of transition. The considered values for n are ½, 2 and 3/2 for allowed transitions, indirect transitions and direct forbidden transitions respectively. CdO is direct energy band gap materials[51]. The extrapolations of the linear plots provide the band gap values for direct allowed transitions. Each of the spectra in figure 8(b) encompasses a broad shoulder in the lower energy side. This broad shoulder is a signature of Cd rich phase. Therefore, it is quite clear that for all the three thin films Cd dominance is present in the relative stoichiometry which also confirmed by RBS results. In Tauc plot it has been assumed that conduction band minima (CBM) and VBM are parabolic in nature and this assumption is not compatible with the actual scenario for CdO. The highly non-parabolic conduction band and carrier concentration dependent band gap can't be determined correctly from Tauc relation [52]. The non-parabolicity of CBM due to existence of additional donor states has a tendency to conceal the absorption spectra near band edge region which one makes unable to calculate the magnitude of indirect band gap value for the concerned material. However this method has been exhaustively used in literature for determination of the band gap.

From the band gap values we observe that for CdO700 thin film band gap has been reduced little bit compared to the CdO500 thin film. This phenomena is trivial as band gap reduction due to increment in temperature is the consequence of electron-phonon coupling and lattice expansion [53]. This phenomena is commonly parameterized by the following equation [54]:

$$E_g(T) = E_g(0) - \frac{\alpha T^2}{\beta + T} \quad (3)$$

where $\alpha$ and $\beta$ are the Varshni parameter, $E_g(T)$ is the band gap value for temperature T K and $E_g(0)$ is the band gap value at 0 K. The values found in the literature for $\alpha = 8*10^{-4}$ eV/K , $\beta = 260$ K and at zero K temperature direct band gap value is 2.31 eV [53]. However putting those values in the equation (3) we find the band gaps for CdO500 and CdO700 thin films are 1.85 and 1.7 eV respectively which are not perfectly matching with our experimental results. The non-parabolicity of the CBM may be the possible reason for this mismatch in the band gap value. For CdO900 thin film, we observe a nontrivial behaviour as band gap has enhanced drastically instead of reducing for the nanosheet like structure. For CdO900 the band gap value enhances to 3.20 eV which is nothing but near to the band gap value of hexagonal wurtzite phase ZnO. Therefore, this enhancement gives a direct signature of structural phase transformation from cubic rocksalt to hexagonal wurtzite phase.

The defects close to the grain surface usually serves a vital role in deciding the optical properties of the material. The grains in the polycrystalline thin films are very much sensitive to their surrounding environment as the surface to volume ratio is very large. The dangling bonds at the interface region between the grains are very much crucial for the presence of localized electronic states which might be responsible for band gap reduction in CdO700 thin film. Therefore, generation of localized energy levels having energy less than the band gap for the crystalline material might be another possible reason for band gap reduction in CdO700 thin film compared to CdO500. From this discussion one can say that even a small fluctuation in the nanocrystallites results in a continuous density of states and also results in a decrement in the energy gap [55]. Whenever the thickness reduces these surface states becomes more important for deciding the band gap [56]. Here grain size doesn't reduce with decreasing film thickness. The whole morphology changes itself and forms a nanosheet like structure. Generally with decreasing film thickness the band gap increases due to quantum confinement phenomena [55,57]. Therefore the band gap enhancement is not a nontrivial phenomenon over here from the perspective of quantum confinement effect. However, it is difficult to explain this band gap enhancement phenomena firmly depending upon quantum confinement effect as thickness is 120 nm which is not very less to fall in the confinement domain and from XRD pattern we can't calculate the particle size for this nanosheet structure. However, the diffuse scattering might come from the rough surface and inhomogeneities of the thin film surface [56] which might help in reducing the band gap by generating some localized electronic energy states. But that is not the scenario in the present case. Therefore this band gap enhancement phenomenon can be described in terms of surface polarizability and dielectric screening effect. It is reported by Delerue *et al.*[58] that the band gap enhancement might be corroborated to the local reduction of polarizability at the surface region. Surface polarization is very sensitive and susceptible to surface condition. For CdO900 thin film we already have found that *rms* roughness is larger. Therefore, in band gap widening phenomena reduction of average dielectric constant and breaking of polarisable bond on the surface take place due to increased roughness [59]. To compensate the uncompensated surface charge the net dipole moment would be liable to diverge and electrostatic potential increases [46]. The increased surface roughness endorses the divergence of net dipole moment and as a consequence polarisable bond breaks which may help in reducing the average dielectric constant and results in a band gap opening. Therefore, rocksalt to wurtzite phase transformation for CdO900 thin film is in agreement with band gap enhancement phenomena.

It is quite well known fact that nanosheet is a bound nanoscopic system [60]. Therefore, the interface geometry which was existing for CdO700 and CdO500 thin film changes completely and the contribution of the surface states to the optical density get reduced and system tries to recover the bulk optical properties of the transformed wurtzite phase [61]. This anticipation is quite compatible with our retrospective discussion in XRD and microscopy section regarding this structural phase transformation.

**III.5. Soft X-ray absorption spectroscopy analysis:**

In figure 9, oxygen *K* edge soft x-ray absorption spectra for CdO500, CdO700, CdO900 thin films are shown which have been recorded in TEY (Total Electron Yield) mode with normalized μ(E) versus photon energy(eV). The prime features a, b beyond the absorption edge i.e. post edge features in A region are mainly generated due to the dipole allowed transition between O 1*s* core states to unoccupied *p* states which resided beyond Fermi level as below Fermi level the electronic states are occupied [62]. If bonding between $O^{2-}$ and $Cd^{2+}$ ions would have been completely ionic then unoccupied *p* states would have been filled. Subsequently, the aforementioned dipole allowed transition would have not taken place, Therefore, bonding between $O^{2-}$ and $Cd^{2+}$ ions is not completely ionic but rather of mixed character [63]. With increasing annealing temperature, hump like feature b, gets vanished for CdO700 thin film. However, in the C region is also getting changed with increasing annealing temperature. It is well reported phenomena that defects like oxygen vacancies plays an important role in deciding overall spectral profile [62]. Native defects of first and second coordination shell influence the spectral profile predominantly. When we are annealing the thin films at higher temperature i.e. for longer time, lattice comes by sufficient time to readjust the atoms and subsequently that will release the stress. As the annealing has been accomplished at oxygen environment, therefore, it will reduce the density of oxygen vacancies inside the lattice. So, the reduction in oxygen vacancies concentration might be the reason behind disappearance of b feature and also changes the overall spectral features in the C region beyond 555 eV which arise due to the interference of multiple scattering signals. For rocksalt IIB-VIA compounds, it is well described observation that strong hybridization takes place within oxygen 2*p* and Cd 4*d* orbitals which might result in positive valence band dispersion from centre of the BZ. Therefore, VBM does not reside at centre of BZ and results an indirect band gap contribution. This indirect band gap is the direct consequence of orbital hybridization with octahedral point symmetry for the rocksalt compound like CdO[64]. The participating electrons in the hybridization i.e. the *d* electrons decides the band gap, cohesive energies, equilibrium lattice parameter etc. [65–67] which means if there is a change in the *d* electronic states, certainly it will suffer modification in the aforementioned properties. When we accomplish derivative of oxygen *k* edge spectra, the inflection point for the absorption edges are found at 539 eV, 538.9 eV, 540 eV for CdO500, CdO700, CdO900 thin films respectively. There is not much difference in the inflection point of CdO500, CdO700 thin films but there is almost 1 eV shift has been observed in CdO900 compared to CdO500 thin film which provides a signature that Fermi level shifting towards higher energy is taking place. This Fermi level pinning is quite compatible with band gap enhancement for CdO900 thin film. As we know, wurtzite phase ZnO is an n type degenerate semiconductor [68] like CdO. For CdO Fermi level ($E_f$) and Fermi stabilization level ($E_{fs}$) both resides above CBM[29]. There is always an intrinsic tendency of $E_f$ for any system to get pinned following $E_{fs}$[69]. As we have observed in Tauc plot that band gap is enhanced with 1 eV for CdO900 thin film. Therefore, this absorption edge shift in oxygen *k* edge spectra can be well correlated with band gap enhancement via Fermi level pinning. As dipole allowed transitions takes place from core occupied to unoccupied states beyond $E_f$. If the whole absorption edge is shifting towards higher energy which means the $E_f$ is also shifting at higher energy. This pinning also provides a direct signature that rocksalt to wurtzite PT is taking place as it is obvious that the position of $E_f$ would be at higher energy beyond conduction band minima for wider band gap wurtzite phase compared to rocksalt phase. Angular momentum projected local density of states provides clear evidence that conduction band minimum substates are constructed of Cd 5*s* states hybridized with O 2*p* states for CdO and this has been well enunciated by Demchenko *et al.* [63]. As $E_f$ pinning is taking place with band gap enhancement, the density of those hybridized electronic states at CBM will also get changed with transforming phase. In the plotted non-normalized data the relative intensity of absorption edge is lower for CdO900 thin film. Reduction of overall film thickness might be the reason behind that which is quite compatible with RBS results.

### III.6. X-ray photoelectron spectroscopy analysis:

Normalized X-ray photoelectron spectroscopic measurements have been performed upon three thin films on surface and after each 10 minutes of etching with 6 keVAr ion bombardment, successively four times. Therefore, total five times XPS spectra are recorded for each of the thin film for O 1*s* and 3*d* with survey scan spectra. The high resolution survey scan spectra are also recorded five times for each thin film which reflects all the existing elements. In figure 10, the survey scan spectra at surface and after fourth etching have been shown. In figure 11, XPS spectra upon surface, after 20 minutes etching and after 40 minutes etching are shown in stacked plot for each thin film. Each of the spectra is fitted with voigt function and Shirley background and further deconvolution of the curves are accomplished for detection of existing phases. All the measurements are referenced with C 1s peak situated at 284.6 eV and peak fitting are performed using CASA software.

In figure 11 (a), For CdO500 thin film, Cd 3d surface spectrum shows three spin orbit doublet which are Cd(i) (CdO, 404.3 eV), Cd(ii) ($CdO_2$, 405.1 eV) and Cd(iii) ($Cd(OH)_2$/$CdCO_3$, 405.9 eV) [70]. $Cd(OH)_2$/$CdCO_3$ is the significant contamination component due to exposure at air which is visible only at the surface not after etching in Cd 3d spectra. In figure 11 (b), the surface O 1s spectra reflects that there is three components i.e. O(i) (CdO, 528.9 eV), O(ii) ($CdO_2$, 529.6 eV), O(iii) ($Cd(OH)_2$/$CdCO_3$, 532.0 eV) [70]. For CdO500 and CdO700 thin film the surface spectra are almost similar except there is a shift of 0.2 eV for Cd 3d spectra towards higher biding energy in CdO700 thin film compared to CdO500 one. O 1s spectra are also in coherence with that of Cd 3d. The values of peak position, area and fwhm are mentioned at table 2 in the supplementary data sheet. The significant contribution for the contamination component gets completely vanished after Ar ion bombardment in Cd 3d spectra but not for O 1s spectra in all the films. For CdO500 and CdO700 thin films contamination phase has been observed at surface but not for CdO900 thin film in Cd 3d spectra. However, that is visible even for CdO900 thin film in O 1s spectrum. Therefore, one can make remark that O 1s spectra are more sensitive for contamination phase compared to Cd 3d. The bombarding Ar ions sputter out the surface components and create localized heating spikes, via deriving energy from the impinging ions. Therefore, decoarbonation of $CdCO_3$ and dehydration of $Cd(OH)_2$ take place. This has been reported by Low and Kamel[71] that $H_2O$ gets lost by decomposing and generating $H_2^+$ and $O^{2-}$ vacancies which further depletes and forms rocksalt phase lattice. In each of the plot after etching Cd 3d and O 1s spectra shift towards higher binding energy compared to the surface ones, with a shift of 0.2 +/-0.05 eV. At surface existing $Cd(OH)_2$/$CdCO_3$ phase is an insulating phase and subsequently it will show a differential charging effect as CdO is a semiconducting phase [72]. After etching, there is a significant reduction of contamination phase as the area under the curve for contamination has been reduced and relatively the area and intensity for both CdO and $CdO_2$ phase has got enhanced which can be attributed to the aforementioned shifting toward higher binding energy. However, the energy difference of 6.75 eV between the doublet i.e. Cd $3d_{5/2}$ and Cd $3d_{3/2}$ remains unaltered. Surface band gap narrowing [73] and surface electron accumulation phenomena [74] are well reported in literature for CdO. In Surface cleaned CdO exhibits surface electron accumulation and downward band bending which subsequently generates sub band states toward higher binding energy which are existing in the bulk [69]. This might be another anticipated reason for such kind of shifting in Cd 3d and O 1s spectra after etching. Low kinetic energy irradiation produces defects like oxygen vacancies inside the lattice [75] in oxides which enhances the carrier concentration [76] and further shifts the Cd 3d and O 1s spectra towards higher binding energy. But if we observe spectra after 40 minutes etching, there is not significant shift of Cd 3d doublet and O 1s singlet compared to the 20 minutes etched spectra. Therefore, decomposition of the surface contamination phases is the prime reason behind this shift and not the irradiation induced defects like oxygen vacancies.

If we watch carefully at Cd 3d and O 1s spectra for CdO900 thin film at surface and for etched ones, there is no signature of $CdO_2$ surface phase [27] which is quite prominent for CdO500 and CdO700 thin films at surface. However, the peak intensity is reduced for CdO700 thin film compared to CdO500 thin film and after each etching for a particular one the same is reduced significantly. This can be clearly observed from the numerical values, derived from fitting for $CdO_2$ phase, mentioned in table 2 of the supplementary data sheet.

There is a thickness dependency of $CdO_2$ surface phase. As CdO900 thin film is having lesser thickness, confirmed from RBS measurements and we do not observe the surface phase. Basically, one can say crudely that we are removing the layers from the thin film surface via annealing with a top down approach. However, there is not any straight forward correlation of rocksalt to wurtzite PT from the Cd 3d and O 1s spectra of CdO900 thin films. Rocksalt to metastable wurtzite PT has been observed by Ashrafi *et al.* for CdO/ZnO heterostructure with increasing thickness [77]. Therefore, thickness plays an important role in such kind of PT. Complete disappearance of $CdO_2$ phase with reduced thickness is the direct experimental evidence for that which is not reported in the literature to the best of our knowledge.

### III.7. Micro-Raman studies

$Fm\bar{3}m$ is the space group symmetry of CdO cubic rocksalt structure. The spectral features for all the three thin films have been shown in figure 12. It is a well-known and universal fact for CdO that A1, E1 both are Raman and IR (Infra Red) active branches, there symmetries are polar, doubly degenerated and split into TO (Transverse Optical) and LO (Longitudinal Optical) components with different frequencies. $E_2(L)$ and $E_2(H)$ are both Raman and infrared active as they are non-polar. In the present investigation Raman analysis have been done in backscattering geometry without taking care about the polarization effect. For CdO500 and CdO700 thin films, we found a sharp peak at 277 cm$^{-1}$ and a less sharp broader hump centred at 946 cm$^{-1}$ also appears. A broad spanning structure from 300 cm$^{-1}$ to 500 cm$^{-1}$ is also observed for both the thin films. For CdO molecule, theoretically it has been reported that LO and TO phonon modes exists at 478(25) cm$^{-1}$ and at 262(3) cm$^{-1}$, respectively [78,79]. There is also a report for existence of 952 cm$^{-1}$ LO phonon mode for CdO molecules. According to the Raman selection rule, both LO and TO modes are dipole forbidden. Therefore all existing features in the Raman spectra are attributed to the second order Raman scattering [80]. During experiment 515.4 nm Ar laser was used which is nearer to the band gap value of the CdO. Therefore, the probability of first order Raman scattering can't be ruled out completely. The polarization vector of the incident light used in the experiment and scattered light are parallel to each other, therefore Raman selection rule would allow the presence of $E_2$ and $A_1$ (LO and TO) modes. The vibrational modes generated from Cd and O sub lattices are related to strain sensitive and non-polar $E_2$(high) and $E_2$(low) modes but independent of the influence of the carriers or the electric field inside the lattice [81]. In CdO700 thin film, we found 5 cm$^{-1}$ red shift in 277 cm$^{-1}$ TO phonon mode as compared to CdO500 thin film. It is reported in the literature that 290 cm$^{-1}$ TO phonon mode is defect incorporated peak for CdO rocksalt phase [77]. Usually after annealing at higher temperature the stress producing defects originated at the time of film preparation start to get annealed out which results in stress relaxation. During annealing lattice gets sufficient time to readjust its atomic positions and we find a 5 cm$^{-1}$ lattice softening for 277 cm$^{-1}$ TO phonon mode. In the retrospective discussion, we observe that a band gap reduction for CdO700 thin film which might reduce the absorption coefficient for both incident and scattered radiation [29]. This may be attributed to the overall narrowing of the Raman peak of CdO700 as compared to the CdO500 thin film. For CdO900 thin film, we found a drastic change in features of spectrum which indirectly indicating the phase transformation. However, it may be noted here that we are not able to match our Raman spectrum exactly with the metastable hexagonal ZnO nanosheet as reported by Umar et al.[20]. Nonetheless, a strong sharp dominant peak has been observed at 437 cm$^{-1}$ for CdO900 thin film.

### III.8 Simulation Results and discussions

The optimized lattice parameter calculated from MD simulation of cubic structure of rocksalt is a=b=c= 4.73 Å and the corresponding bond length of Cd-O is 2.35 Å which is in good agreement with literature[82–85]. At 900 °C, we have observed that the rocksalt structure changes to hexagonal wurtzite phase as shown in figure 13. The corresponding optimized lattice parameters for hexagonal wurtzite phase are a=3.60 Å and c/a ratio is 1.55.However, the bond length is getting slightly reduced in Cd-O to 2.25 Å, also matches with previous reported work[82,84,85].

The calculated electronic band structures of CdO for both the phases are depicted in Figure 14. The electronic band structure of rocksalt shows indirect band gap of 1.06 eV from L to Γ points while the direct band gap of 1.15 eV at 500 °C (Figure 14(a)). Through the molecular dynamics simulation at 700 °C, the electronic structure remains same phase and the calculated electronic band structures shows reduced band gap. The corresponding indirect band gap from L to Γ points is 0.88 eV and direct band gap is 1.06 eV (Figure 14(b)). The band gap of rocksalt is in excellent agreement with previously reported experimental as well as theoretical reports [82–85] i.e. it varies from 0.84 eV to 2.0 eV for indirect band gap and 2.28 eV to 2.68 eV for direct band gap. At 900 °C, the electronic structure obtained from molecular dynamics simulation reflects that it has wurtzite phase as represented in Figure 13. The corresponding electronic band structure shown in figure 14(c) gives direct energy band gap of 1.24 eV at Γ point and very high indirect band gap of 4.02 eV from Γ to A points. It is commonly observed in DFT calculations that the energy band gap values gets underestimated asthe excited states are kept at lower energy and consequently the conduction band minima also resides at lower energy [28]. To better understand the contribution of electronic states, we have also calculated the projected density of states. A phase transition i.e. from rocksalt to wurtzite phase in the CdO system with increasing annealing temperature subsequently changes the symmetry, nearest neighbour bond length and coordination number. Consequently, there is a significant impact of aforementioned changes in the band repulsion and hybridization between O 2p and Cd 4d orbitals [28]. The main contribution in the vicinity of Fermi level is generated from p states of O atom and d states of Cd atom in the valence band maximum (VBM). Near the Fermi level, p states of O atom are more dominant as compared to d sates of Cd which is presented in figure 14 (a,b). While in case of wurtzite structure of CdO, the p states of O and d states of Cd both are highly dominant in VBM as shown in Figure 14(c). In wurtzite phase, the CBM level gets shifted towards the higher energy as compared to rocksalt phase. As we know the coordination number is 8 and 4 in case of in rocksalt and wurtzite respectively. Each oxygen atom in rocksalt structure is surrounded by 8 cadmium atoms (transition metals) and has sufficient number of electrons to share in bonding with oxygen. That's why some electrons gets bonded to Oxygen atom and some electrons are free. Due to the presence of sufficient number of free electrons, rocksalt phase have some electronic states near the Fermi level in CBM which is responsible for reducing the band gap as compared to wurtzite structure. Also, the repulsive interaction between Cd 4d and O 2p orbital results a difference in band width between rocksalt and wurtzite structure. The octahedrally coordinated rocksalt structure has inversion symmetry whereas it is absent in tetrahedrally coordinated wurtzite phase. As we know, inversion symmetry means that for every point inside the lattice (x, y, z), there would be an indistinguishable corresponding point at (-x, -y, -z). Because of this inversion symmetry, the repulsive interaction between p and d orbital is suppressed at the centre of the BZ i.e. at Γ point and O 2p bands are pushed upward at the zone boundary region. But it is not possible in the case for wurtzite phase, as repulsive interaction between orbitals resides throughout the BZ. This might be also the possible reason why VBM resides at zone boundary for rocksalt phase and at zone centre for wurtzite phase in the electronic band structure [28].

**IV. CONCLUSIONS**

Temperature dependent rocksalt to wurtzite structural PT has been thoroughly scrutinized in the present report for CdO thin films which has been reported to be hydrostatic pressure driven in most of the cases in the literature. Microscopy studies not only supports the nanosheet formation but also approves PT from TEM images with matching inter planner spacing value with XRD for (101) plane of wurtzite phase for CdO900 thin film. Formation of nanosheet has been explained in the framework of hexagonally transformed wurtzite phase where Cd rich chemically active (0001) polar planes energetically favour the subsequent adsorption and nucleate such chemically active planes throughout the substrate and mingle together due to long range electrostatic interaction among the polar charges of (0001) planes.The enhanced surface roughness, calculated from AFM images diverges the net dipole moment which might break the polarizability bonds and also might help in reducing the average dielectric constant and results in a band gap opening. Further, Oxygen *k* edge analysis approves the band gap enhancement with Fermi level pinning. XPS measurements reflect the presence of contaminations phases at thin film surface and also provide a direct experimental signature of thickness dependent $CdO_2$ phase. MD simulation and band structure calculation with DFT approves the PT theoretically. Thus, in the present manuscript combined experimental and theoretical investigation have been accomplished regarding PT and its deep correlation with nanosheet network formation which might have a significant ramification not only towards physics fraternity but also from optoelectronic applications point of view.

## Acknowledgments

Authors are grateful to the Director (IUAC) for his encouragement and moral support for extending experimental facilities. Authors are also grateful Mr. G. R. Umapathy, Dr. Indra Sulania, Dr. Saif Ahmend Khan, Dr. D.K. Shukla and Dr. Fouran Singh for experimental support in RBS, AFM, SEM, SXA investigations and discussions respectively. One of the authors (Dr. A. Das) acknowledges University Grant Commission, New Delhi for Senior Research Fellowship (SRF) Grant Number-F.2-91/1998(SA-1). DST, Govt. of India is also gratefully acknowledged for providing FE-SEM through nano-mission project and SERB for granting project (SB/EMEQ-122/2013).

**Table and Figure captions**

**Table 1:** Numerical values obtained from the different characterizations

**Figure 1:** (a) Experimentally observed Rutherford backscattering spectra for CdO500, CdO700, CdO900 thin films whereas their corresponding simulated results are shown in (b), (c) and (d), respectively.

**Figure 2:** Grazing angle X-Ray diffraction patterns for CdO500, CdO700, CdO900 thin films.

**Figure 3:** Plane view scanning electron microscopy image for (a) CdO500, (b) CdO700 and (c) CdO900 thin films, whereas their corresponding magnified view are shown in (a'),(b') and (c'), respectively.

**Figure 4:** Plane view scanning electron microscopy image and corresponding electron dispersive X-Ray spectroscopy image for CdO900 thin film measured at (a) dark and (a') light region, respectively.

**Figure 5:** Atomic force microscopy images for CdO500, CdO700, CdO900 thin films in 5 μm and 2 μm scale.

**Figure 6:** (a) Atomic force microscopy and corresponding (b) scanning electron microscopy image for CdO900 thin film.

**Figure 7:** Transmission Electron Microscopy images for CdO500 and CdO900 thin film

**Figure 8:** (a) Transmittance spectra for CdO500, CdO700, CdO900 thin films whereas corresponding Tauc plot for band gap calculation are shown in (b), respectively

**Figure 9:** O *K*-edge spectra for CdO500, CdO700 and CdO900 thin films

**Figure 10:** Survey scan spectra for CdO500, CdO700 and CdO900 thin films

**Figure 11:** X-ray photoelectron spectroscopy spectra forCdO500, CdO700 and CdO900 thin films

**Figure 12**: Raman spectra for CdO500, CdO700, CdO900 thin films

**Figure 13**: Full optimized structures of rocksalt and wurtzite phase

**Figure 14**: The electronic band structure and corresponding projected density of states of (a) rocksalt at 500 °C, (b) rocksalt at 700 °C and (c) hexagonal wurtzite at 900 °C.

**Table- 1**

| Sample name | Particle diameter (nm) | Rms roughness (nm) | Band gap (eV) | Film thickness (nm) | Cd composition from RBS | O composition from RBS |
|---|---|---|---|---|---|---|
| CdO500 | 27 | 14.5 | 2.22 | 180 | 85% | 15% |
| CdO700 | 31 | 5.9 | 2.12 | 170 | 74% | 26% |
| CdO900 | n.a. | 27 | 3.20 | 120 | 20% | 80% |

*n.a. mean "not applicable"*

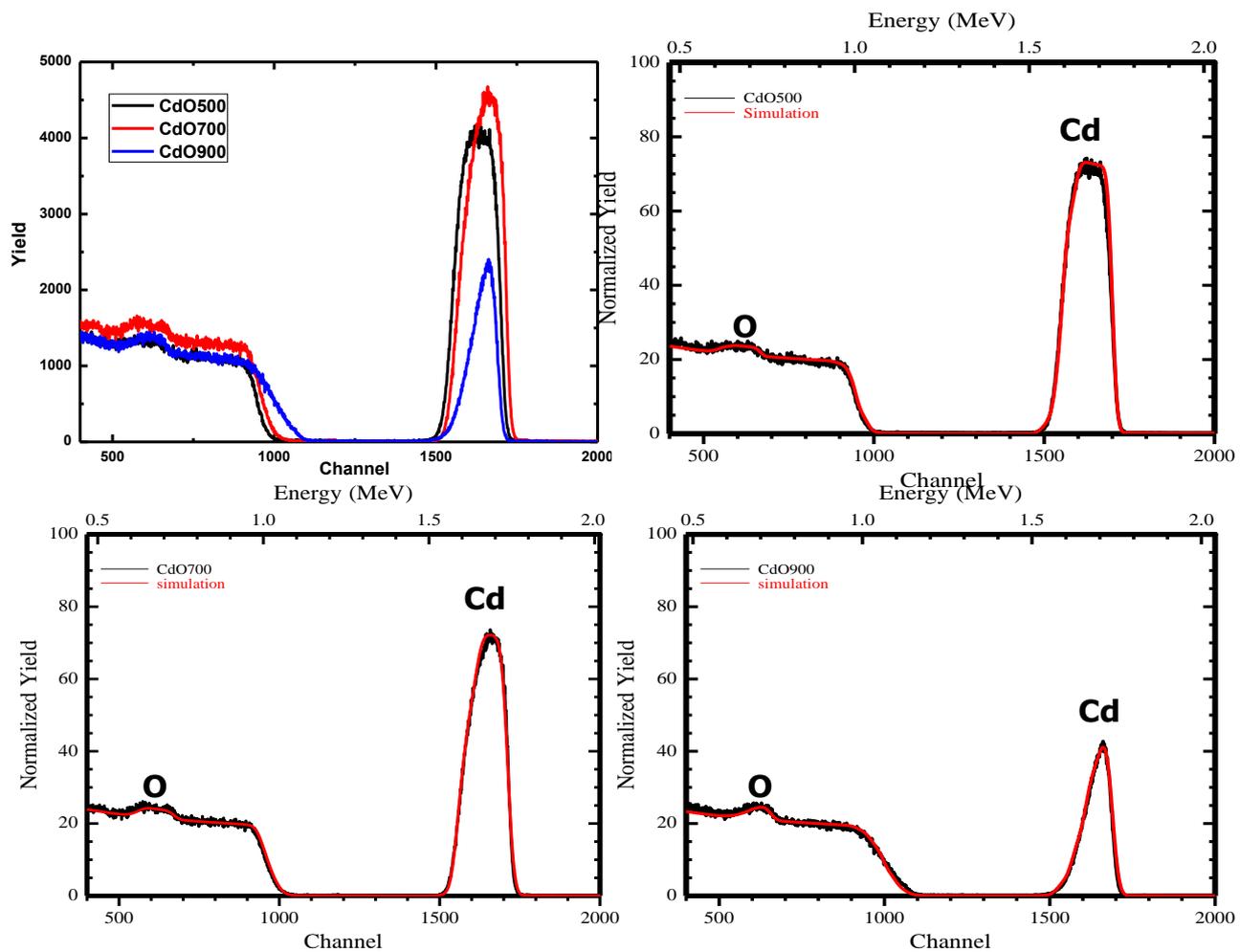

**Figure 1**

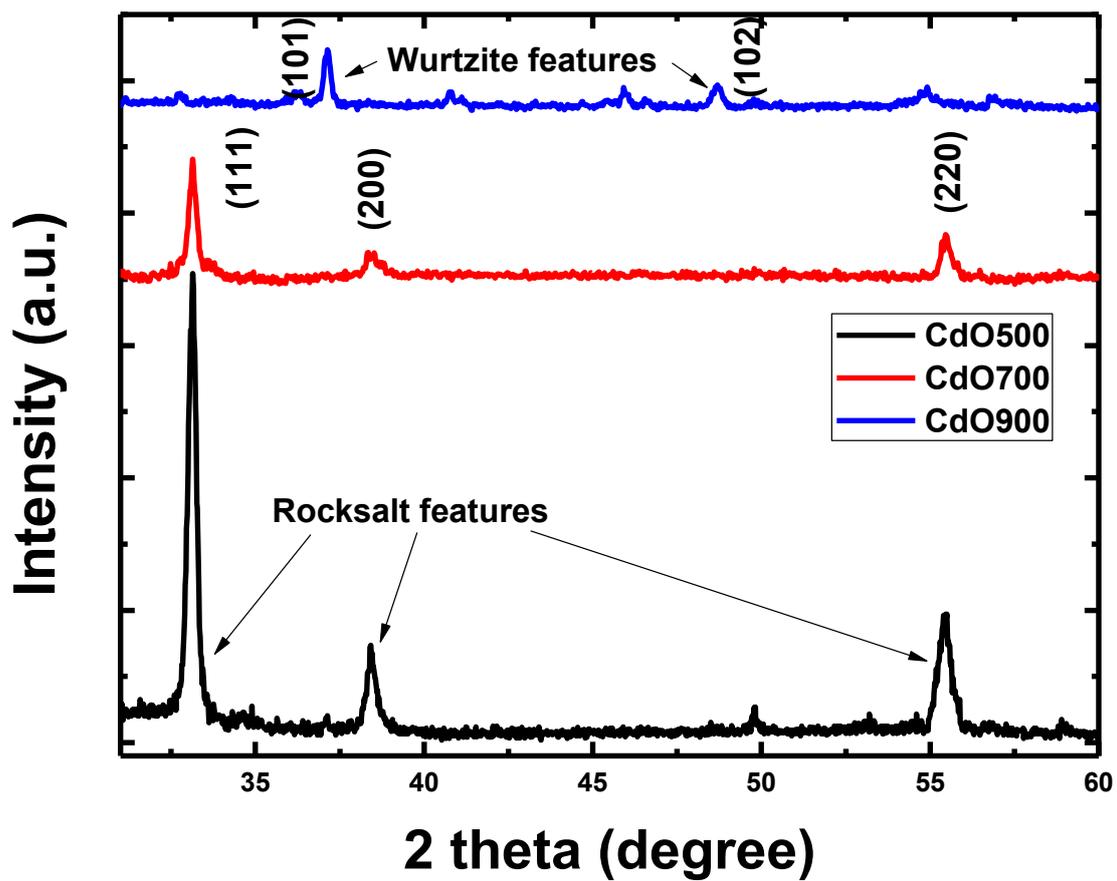

Figure 2

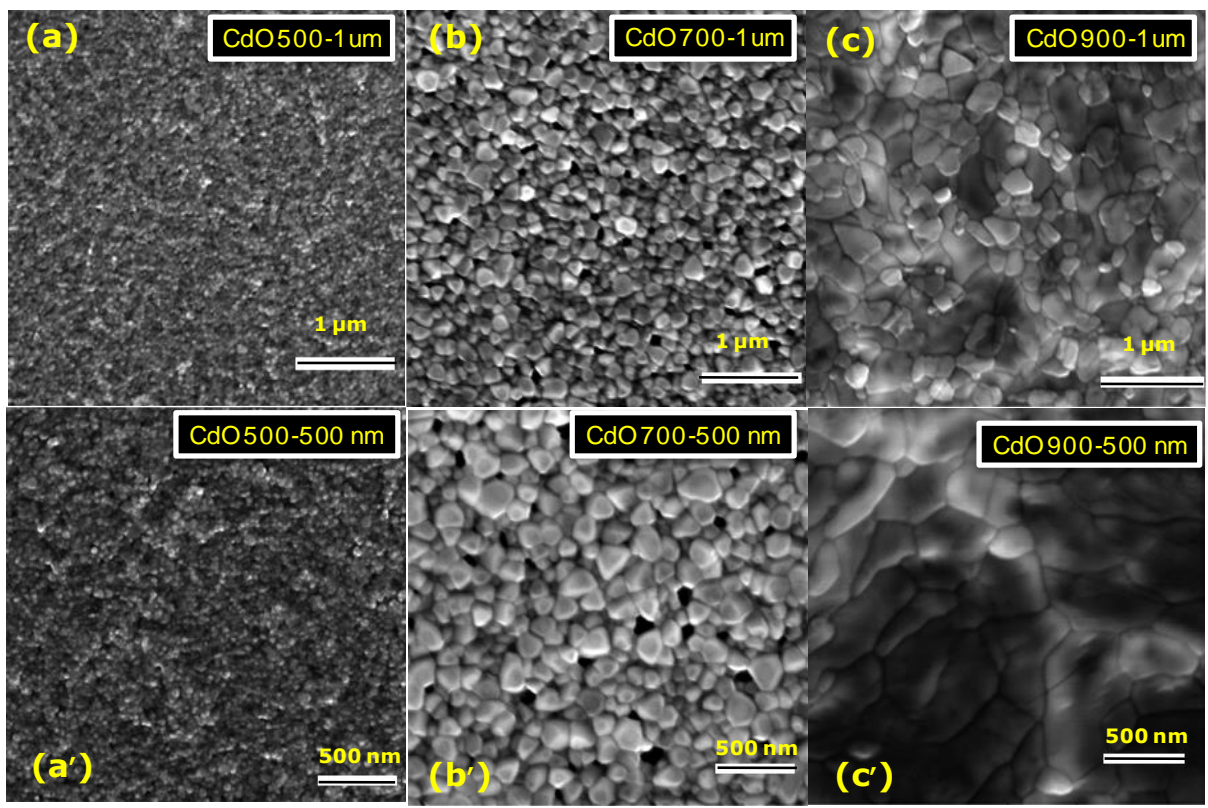

Figure 3

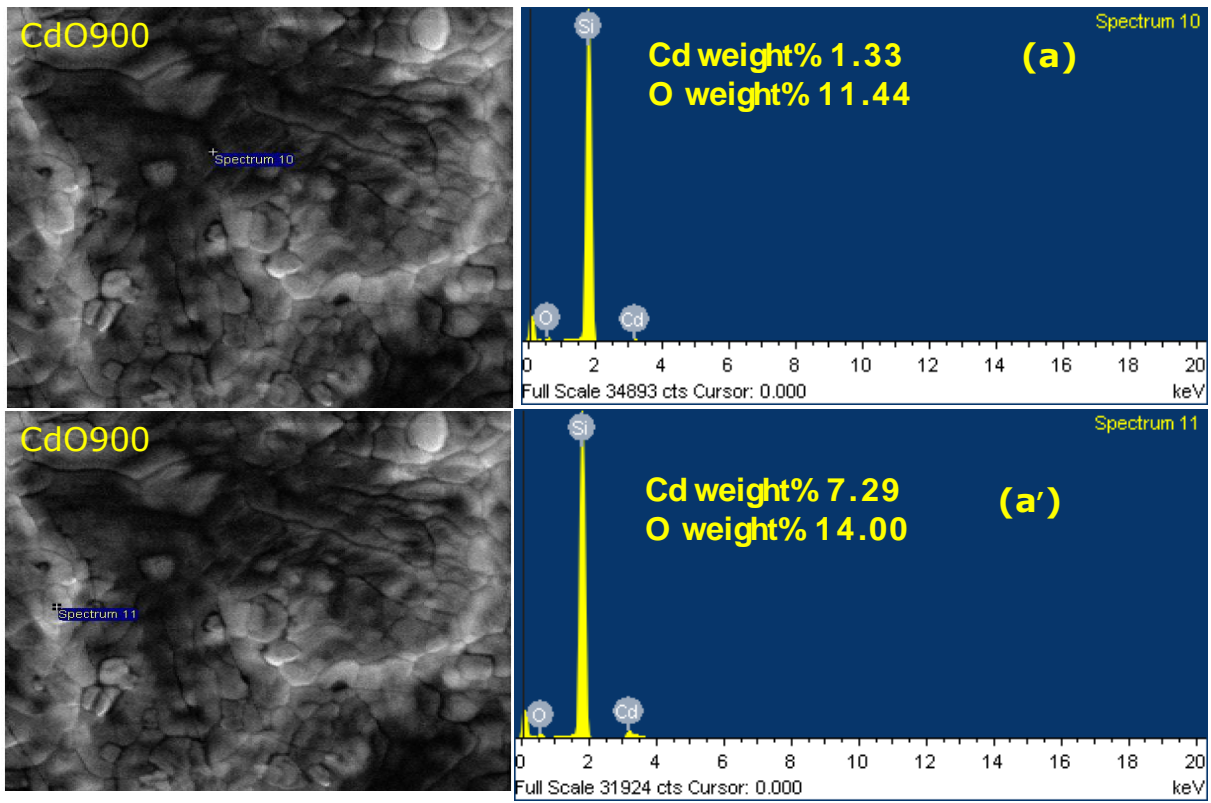

**Figure 4**

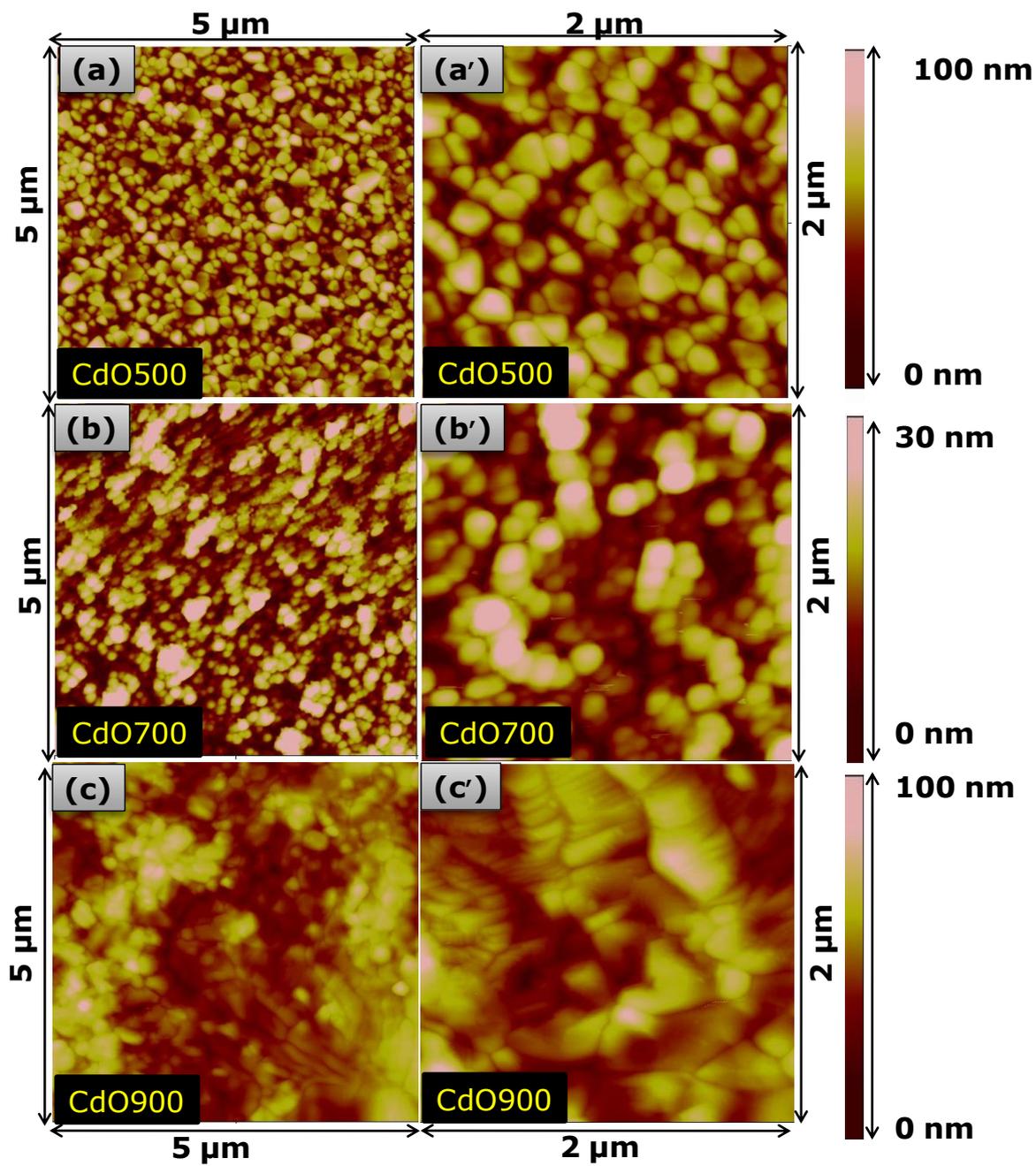

**Figure 5**

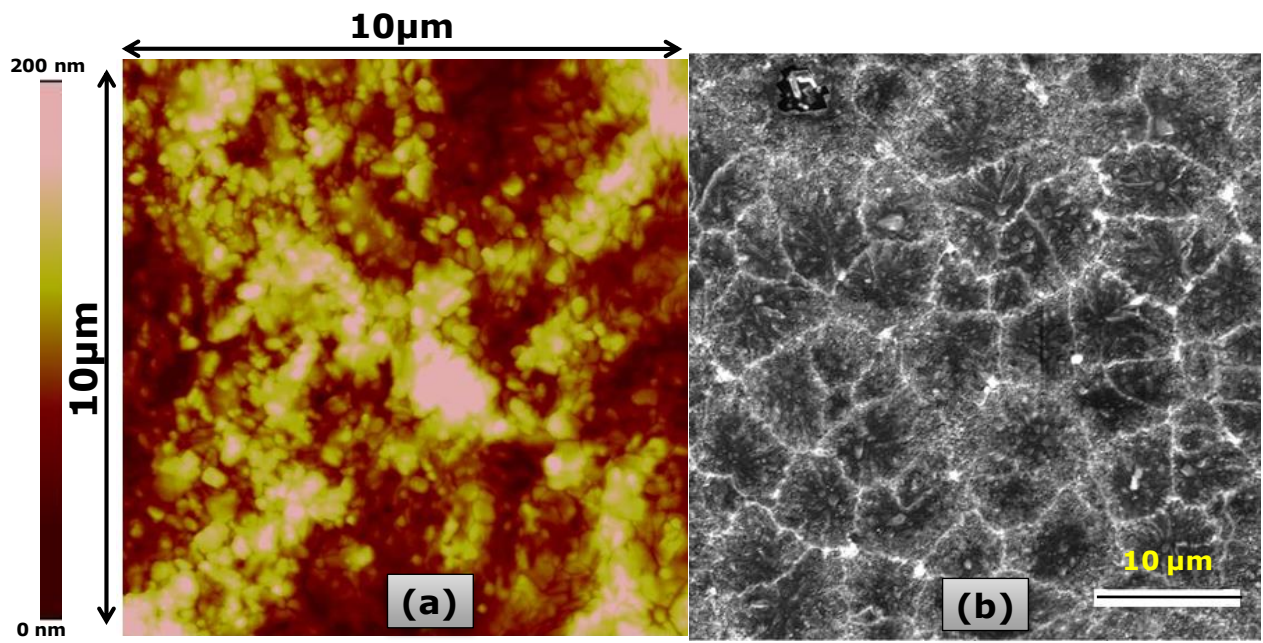

**Figure 6**

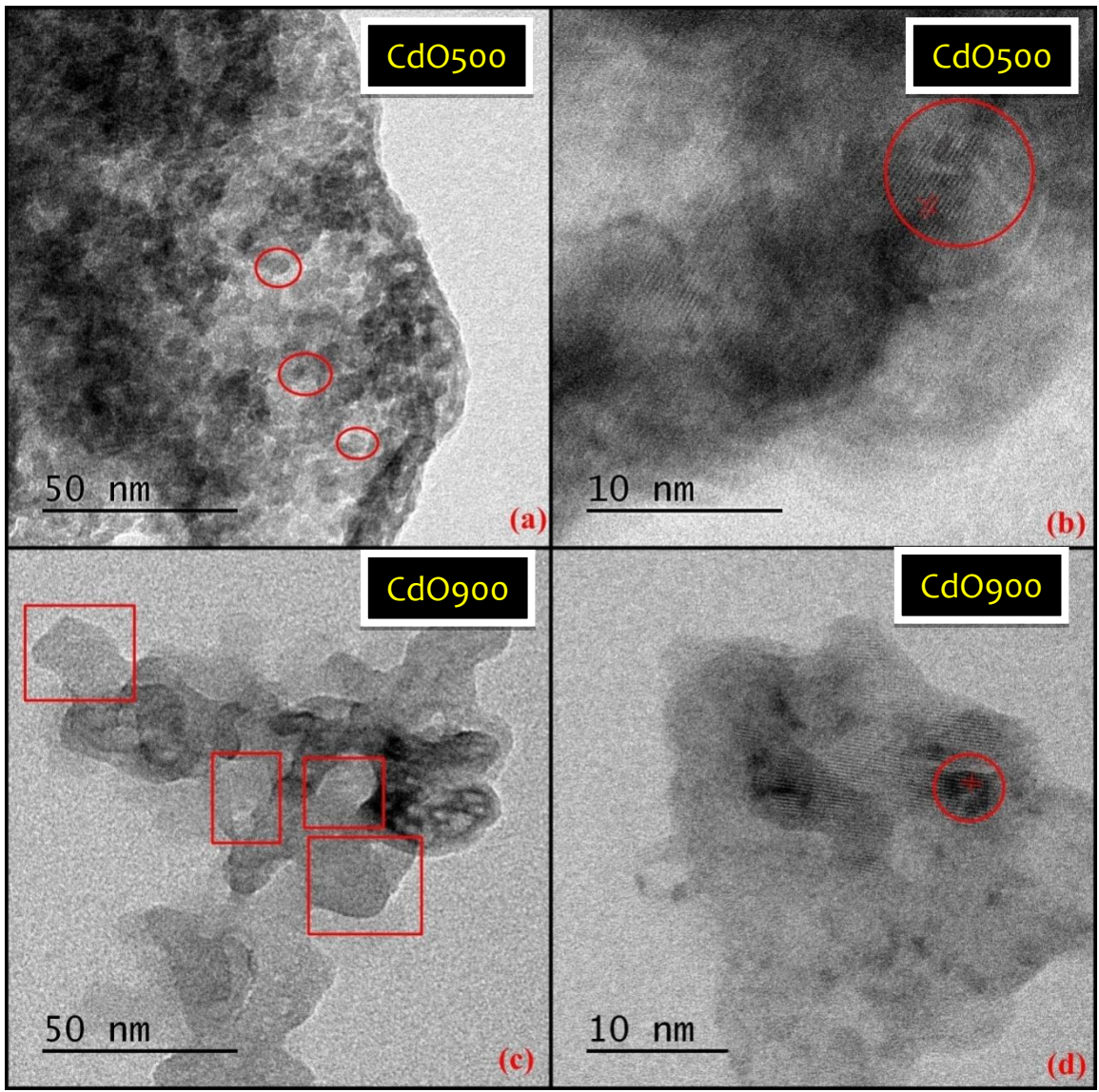

**Figure 7**

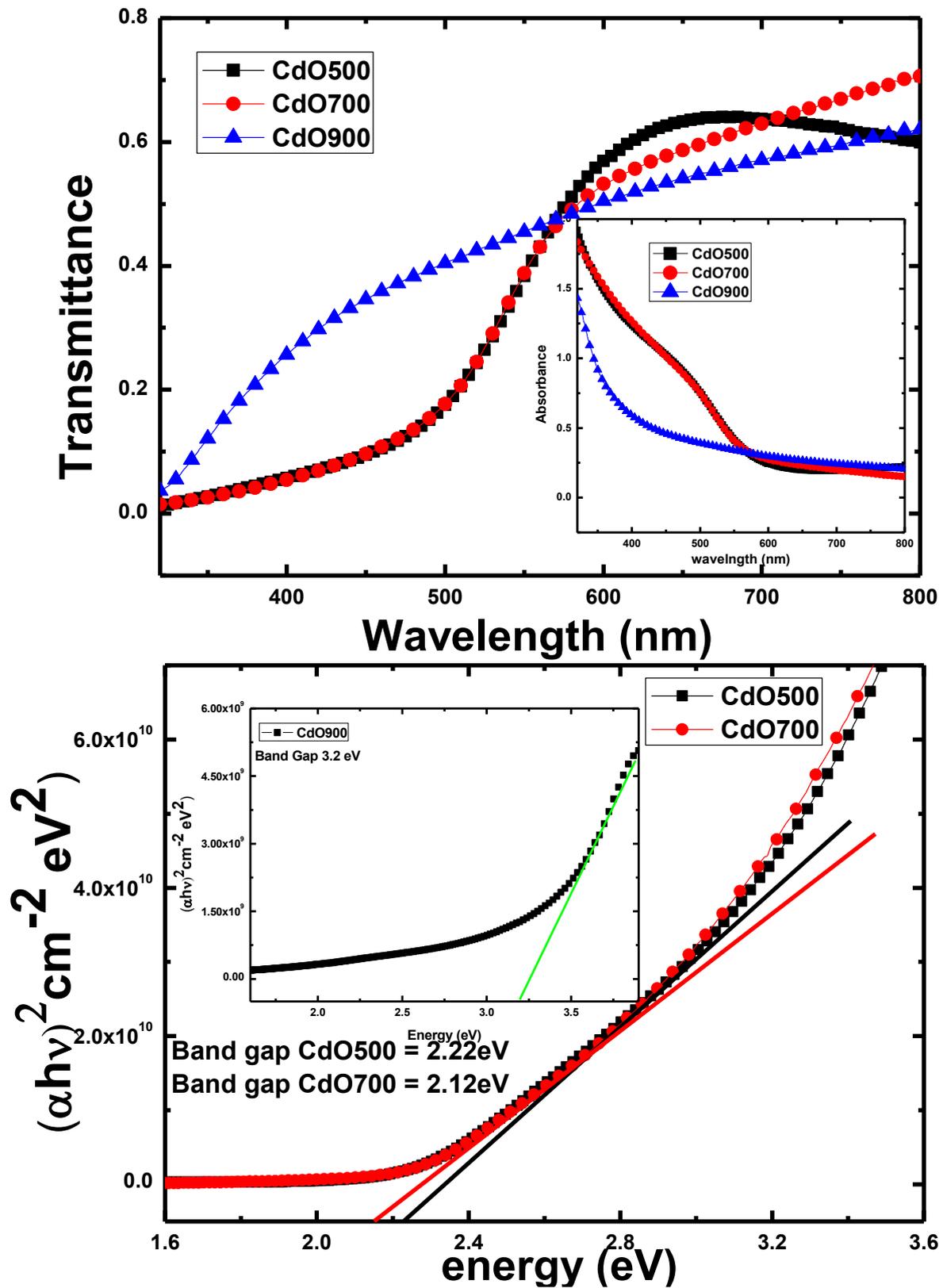

**Figure 8**

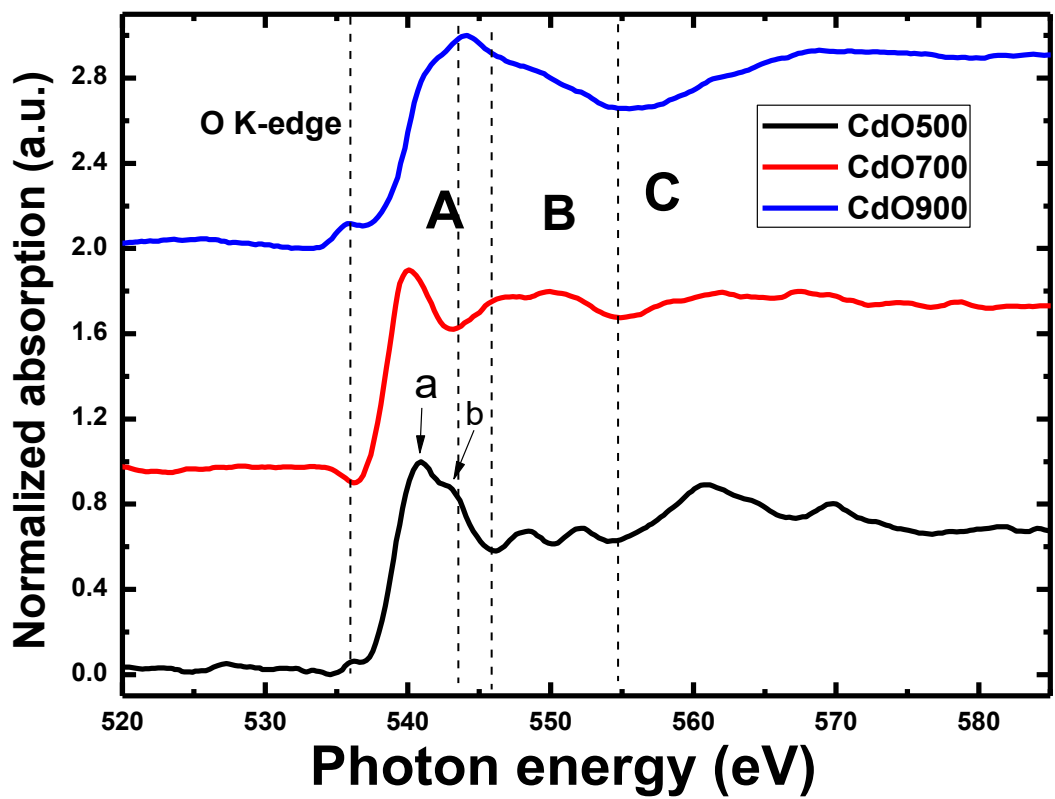

Figure 9

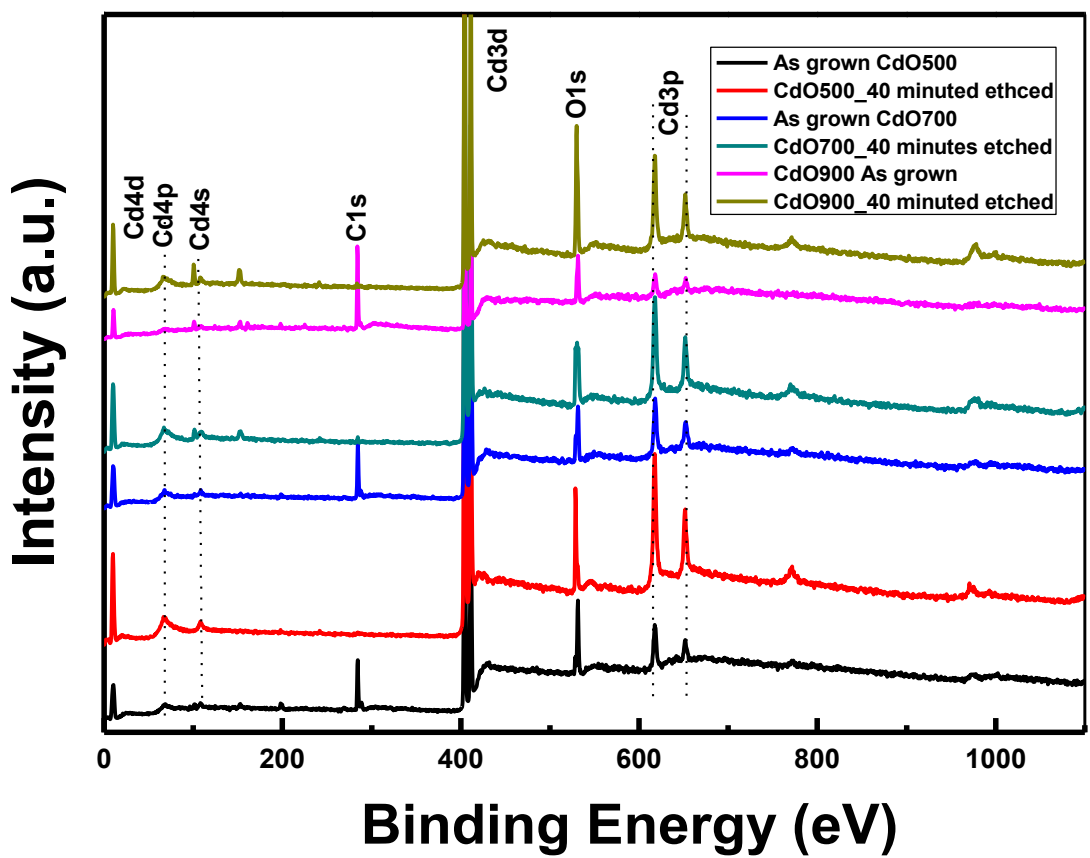

Figure -10

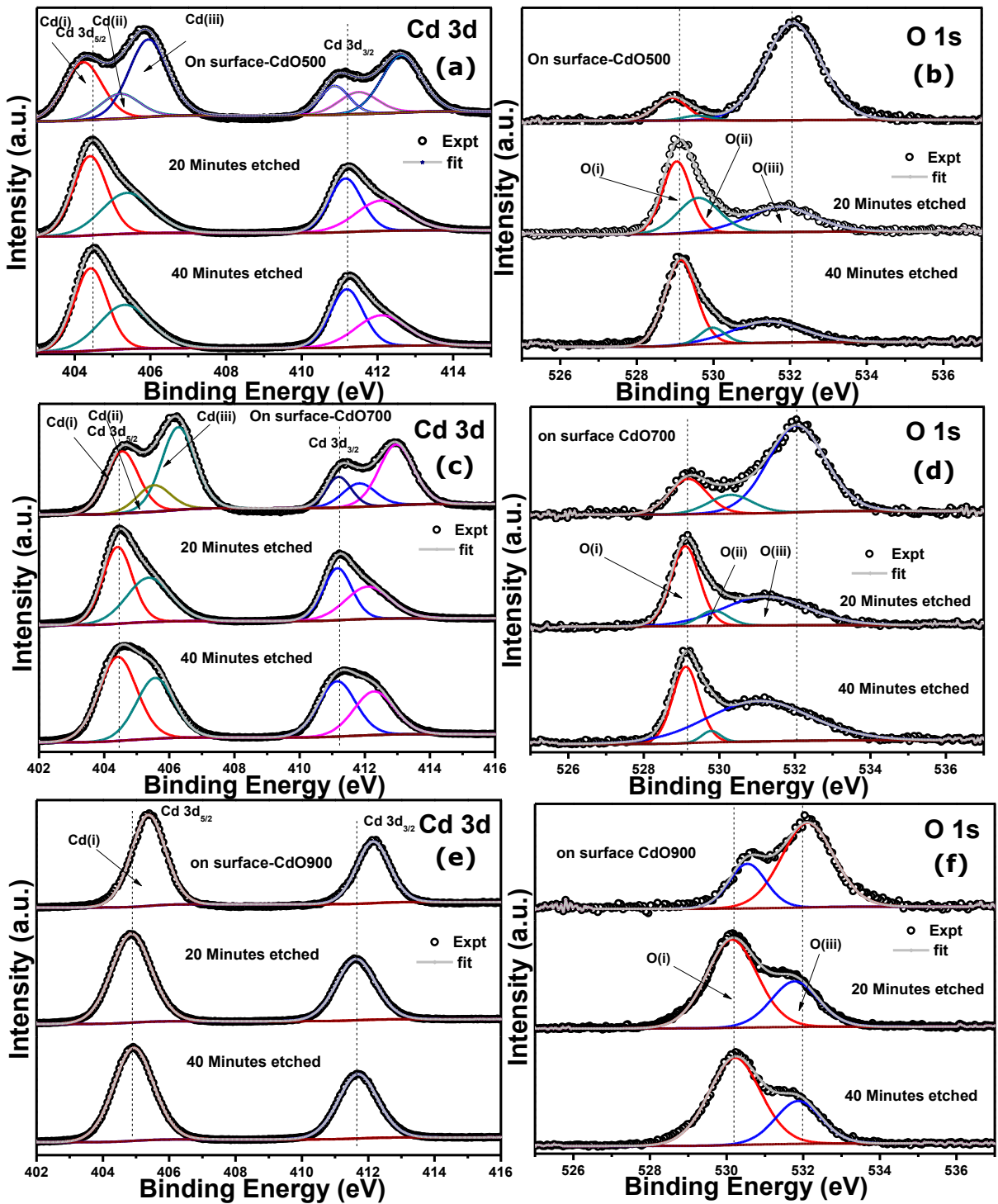

**Figure -11**

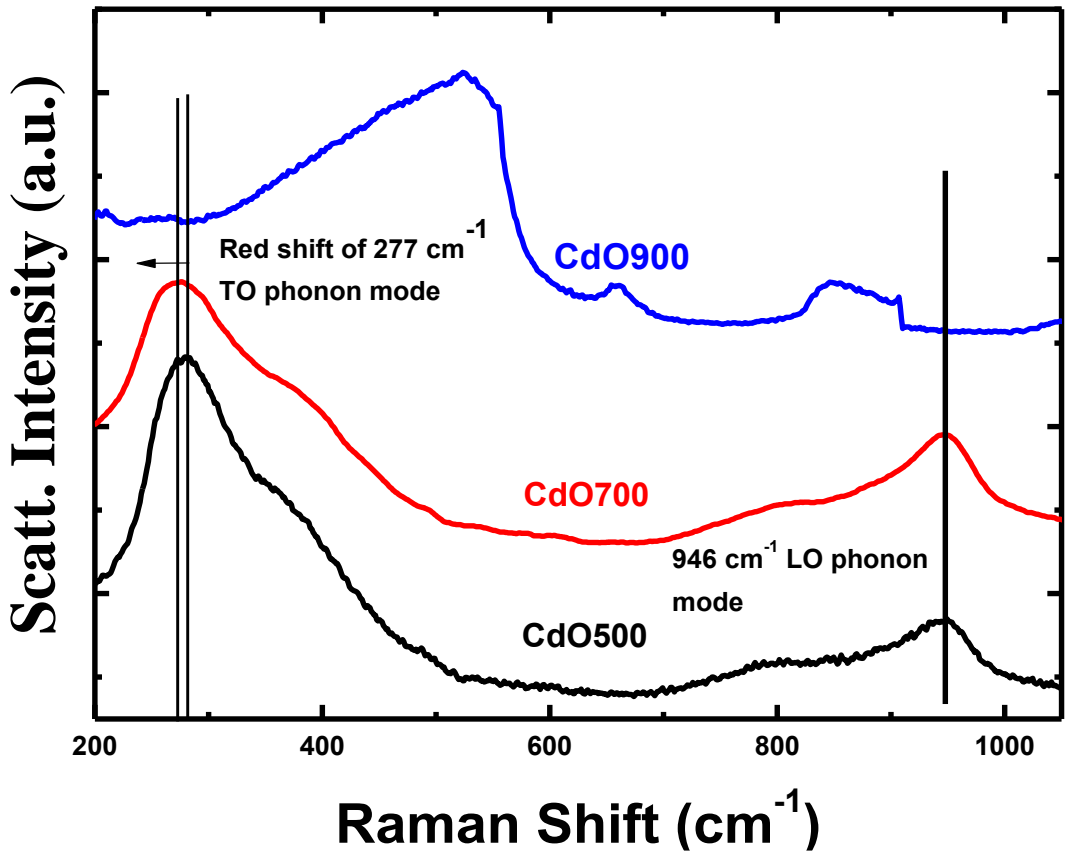

**Figure -12**

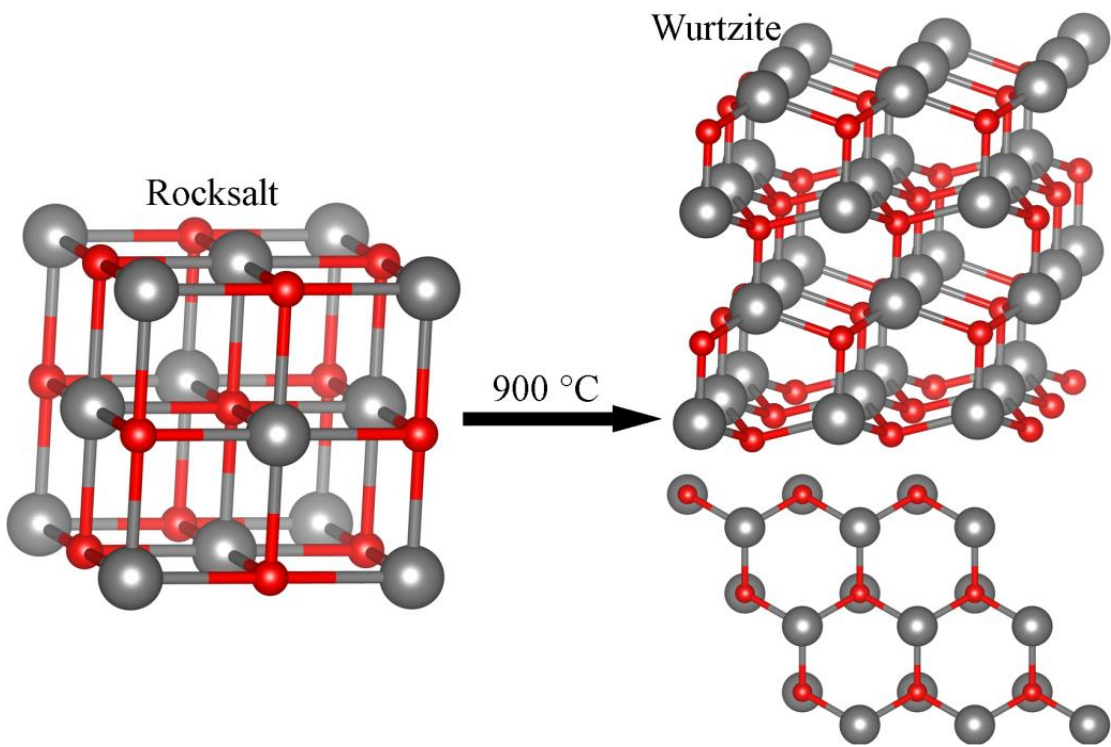

**Figure 13**

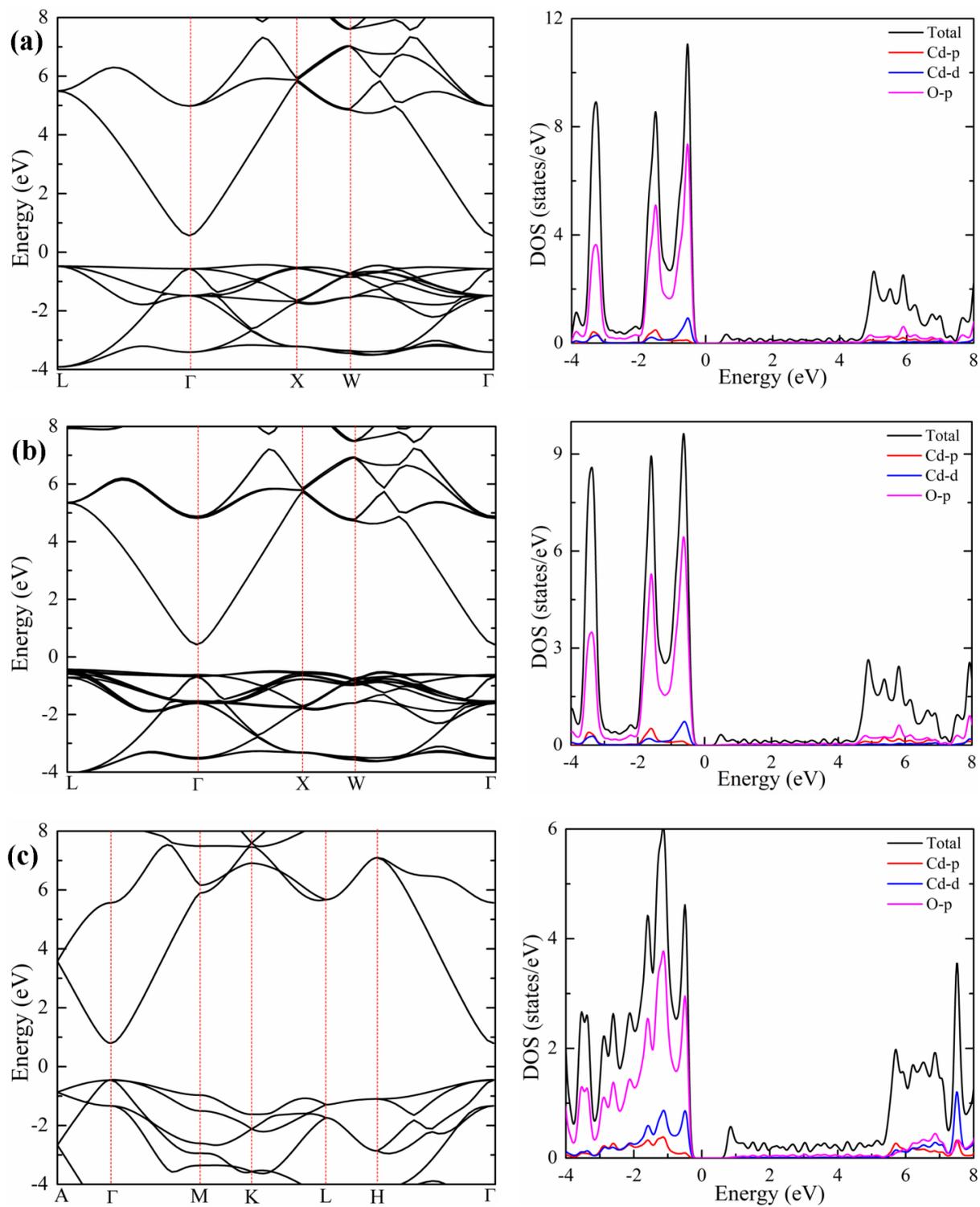

**Figure-14**